\documentclass[aps,prd,amssymb,amsmath,eqsecnum,nofootinbib,twocolumn]{revtex4}
\usepackage{xcolor,ulem}
\usepackage{scalerel}
\usepackage{algorithm}
\usepackage{algpseudocode}
\usepackage[a4paper,total={7in,9in}]{geometry}

\usepackage{color}
\usepackage{bm}
\usepackage{ulem}
\usepackage{graphicx}
\graphicspath{{./img/}}
\usepackage{scalerel}
\usepackage{tikz}
\usetikzlibrary{svg.path}

\definecolor{orcidlogocol}{HTML}{A6CE39}
\tikzset{
  orcidlogo/.pic={
    \fill[orcidlogocol] svg{M256,128c0,70.7-57.3,128-128,128C57.3,256,0,198.7,0,128C0,57.3,57.3,0,128,0C198.7,0,256,57.3,256,128z};
    \fill[white] svg{M86.3,186.2H70.9V79.1h15.4v48.4V186.2z}
                 svg{M108.9,79.1h41.6c39.6,0,57,28.3,57,53.6c0,27.5-21.5,53.6-56.8,53.6h-41.8V79.1z M124.3,172.4h24.5c34.9,0,42.9-26.5,42.9-39.7c0-21.5-13.7-39.7-43.7-39.7h-23.7V172.4z}
                 svg{M88.7,56.8c0,5.5-4.5,10.1-10.1,10.1c-5.6,0-10.1-4.6-10.1-10.1c0-5.6,4.5-10.1,10.1-10.1C84.2,46.7,88.7,51.3,88.7,56.8z};
  }
}

\newcommand\orcidicon[1]{\href{https://orcid.org/#1}{\mbox{\scalerel*{
\begin{tikzpicture}[yscale=-1,transform shape]
\pic{orcidlogo};
\end{tikzpicture}
}{|}}}}

\usepackage{hyperref}

\begin{document}

\title{Apparent horizon tracking in supercritical solutions of the Einstein-scalar field equations in spherical symmetry in affine-null coordinates}

\author{Thomas M\"adler \orcidicon{0000-0001-5076-3362}}
\email{thomas.maedler._.at._.mail.udp.cl}
\affiliation{Escuela de Obras Civiles and Instituto de Estudios Astrof\'isicos, Facultad de Ingenier\'ia y Ciencias, Universidad Diego Portales, Av. Ej\'ercito Libertador 441, Santiago, Chile }

\author{Olaf Baake \orcidicon{0000-0003-4353-1058}}
\email{olaf.baake._.at._.inst-mat.utalca.cl}
\affiliation{Instituto de Matem\'aticas (INSTMAT), Universidad de Talca, Casilla 747, Talca, Chile}

\author{Hamideh Hosseini \orcidicon{0009-0003-4073-1828}}
\affiliation{ Instituto de Estudios Astrof\'isicos, Facultad de Ingenier\'ia y Ciencias, Universidad Diego Portales, Av. Ej\'ercito Libertador 441, Santiago, Chile }

\author{Jeffrey Winicour \orcidicon{0000-0003-0902-1951}}
\affiliation{Department of Physics and Astronomy, University of Pittsburgh, Pittsburgh, Pennsylvania 15260, USA}

\begin{abstract}

Choptuik's critical phenomena in general relativity is revisited in the affine-null metric
formulation of Einstein's equations for a massless scalar field in spherical symmetry. Numerical solutions
are obtained by evolution of initial data using pseudo-spectral methods. The underlying
system consists of differential equations along
the outgoing null rays which can be solved in sequential form.
A new two-parameter family of initial data is presented for which these equations can be integrated analytically. Specific choices of the initial data parameters correspond to either
an asymptotically flat null cone, a black hole event horizon or the singular interior of a black hole.
Our main focus is on the interior features of a black hole, for which the affine-null system
is especially well adapted. 
We present both analytic and numerical results describing the geometric
properties of the apparent horizon  and final singularity.
Using a re-gridding technique for the affine parameter, numerical evolution of initially
asymptotically flat supercritical data can be continued inside the event horizon and track
the apparent horizon up to the formation of the final singularity.

\end{abstract}

\maketitle

\section{Introduction}

Choptuik's  \cite{Choptuik:1993PRL} discovery of  critical phenomena in general relativity  is one  of the first  major
results of the numerical investigation of the Einstein equations. At its heart is the
discretely self-similar (DSS)  critical solution of the Einstein equations for a 
spherical symmetric massless-scalar field 
which evolves to form a naked singularity with zero mass. The scalar field is periodic
with respect to a time scale adapted to the discrete conformal symmetry, with twice the echoing period
$\Delta\approx 3.44$
of the corresponding conformal metric \cite{Choptuik:1993PRL, 
1995PhRvL..75.3214G,1997PhRvD..55.3485H,1996CQGra..13..497H}. Until recently,
the existence of the critical solution and its properties have
only been inferred by numerical evolution.  However,  the existence of this critical DSS solution has been established
and its properties studied by purely
analytic methods \cite{2019CMaPh.368..143R}. 
By calculating the inverse of an elliptic operator, the authors of  \cite{2019CMaPh.368..143R} provide a value for $\Delta$ with an accuracy of $10^{-80}$.

Choptuik observed that a one parameter family of asymptotically flat initial data, with parameter $p$,  evolves to either a flat space-time or to a black hole, with the two alternatives
intermediated  by a critical value $p^*$. Here $p< p^*$ for subcritical
(weak) initial data and $p> p^*$, for supercritical (strong) initial data.
For supercritical evolution, he found that the mass $m$ of the  black hole obeys a universal scaling relation $m\sim |p-p^*|^\gamma$, with $\gamma\approx0.37$, independent of the particular
form of the initial data. 
{Further analysis \cite{1997PhRvD..55..695G,Hod:1996ar} revealed a modified scaling law $\ln(m) =\gamma \ln|p^*-p| + F\left[\frac{2\pi}{\Delta}(p-p^*)\right]$, where $F$ is a periodic function of $\ln(p-p^* )$ 
with period $(\Delta/2\gamma)\approx 4.61$}.
(See \cite{2007LRR....10....5G} for a review.)

For the critical case $p=p^*$, there is a value $\tau=\tau^*$ of the proper time along the central geodesic for formation of the final singularity,
with  $\tau^* $ dependent upon the particular initial data. 
For supercritical initial data, 
{the evolution leads to a black hole for
$p > p^*$.}
Several studies, e.g. \cite{1997PhRvD..55.3485H,1996CQGra..13..497H,2003PhRvD..68b4011M},  indicate that the echoing period $\Delta$ is adapted to
the ``logarithmic time" coordinate $\xi = -\log (\tau^*-\tau)$,
i.e. the metric satisfies $ e^{-2\Delta}g_{ab}(\xi+\Delta)  =g_{ab}(\xi) $
and the scalar field satisfies $\Phi(\xi+ \Delta) =- \Phi(\xi)$.

{The resolution of the critical behavior is numerically challenging. Because  the echoing occurs with respect to the logarithmic time $\xi$, the time $\Delta \tau$ between subsequent periods decreases exponentially
on approach to $\tau^*$. In addition, the spatial structure appears on ever smaller scales. As a result, a fixed spatial grid eventually
does not include enough points near the central worldline  to resolve the region of interest \cite{Garfinkle:1994jb,Purrer:2004nq}. }

In this work, we extend the investigation \cite{CW2019} of critical collapse in affine null
coordinates $(u,\lambda, x^A)$, where $u$ measures the proper time along the central
timelike geodesic and $\lambda$ is an affine parameter along the spherical congruence of outgoing null rays, which are labeled by angular coordinates $x^A$. We present the underlying
formalism in Sec. \ref{sec:2}. 
We modify the evolution algorithm in  \cite{CW2019} to
allow tracking of the apparent horizon.
{See \cite{Csizmadia:2009dm} for a Cauchy evolution study with comparable tracking
of the apparent horizon.} 
We also  introduce a novel 
two-parameter set of initial data for which the null hypersurface equations can be 
integrated  analytically. Specific choices of the initial data parameters correspond to either
an asymptotically flat null cone, a black hole event horizon or the singular interior of a black hole.

As opposed to the numerical investigation of the Choptuik problem in  \cite{2002PhRvD..65h1501L,Purrer:2004nq}
using Bondi null coordinates based upon a surface area radius $r$, the affine null system
allows evolution inside the event horizon where the $r$ coordinate is singular at the apparent horizon. {By contrast, the evolution
inside the event horizon breaks down in Bondi coordinates} (see \cite{Win2013} for further discussion). For the choice of black hole initial data, the affinely parametrized
null cones extend smoothly across the apparent horizon, where the expansion of
the outgoing null cones vanishes, and up to the final singularity, where the null cones re-converge to a point. This  allows a combination of analytic and numerical methods to investigate the interior of the black hole.

In comparsion with  \cite{CW2019}, we use a single domain spectral
method based on the standard Chebysheff polynomials, combined with the grid
compactification of null infinity described in \cite{2012LRR....15....2W} and
used  in several other characteristic codes
\cite{2015CQGra..32b5008H,2016CQGra..33v5007H}.
Accuracy near the central worldline is increased by filling 
a local set of collocation points with
values obtained from a Taylor series about the origin.
Further details of the numerical techniques are given in Sec.~\ref{sec:num_detail}.

\begin{figure}
\includegraphics[scale=0.4]{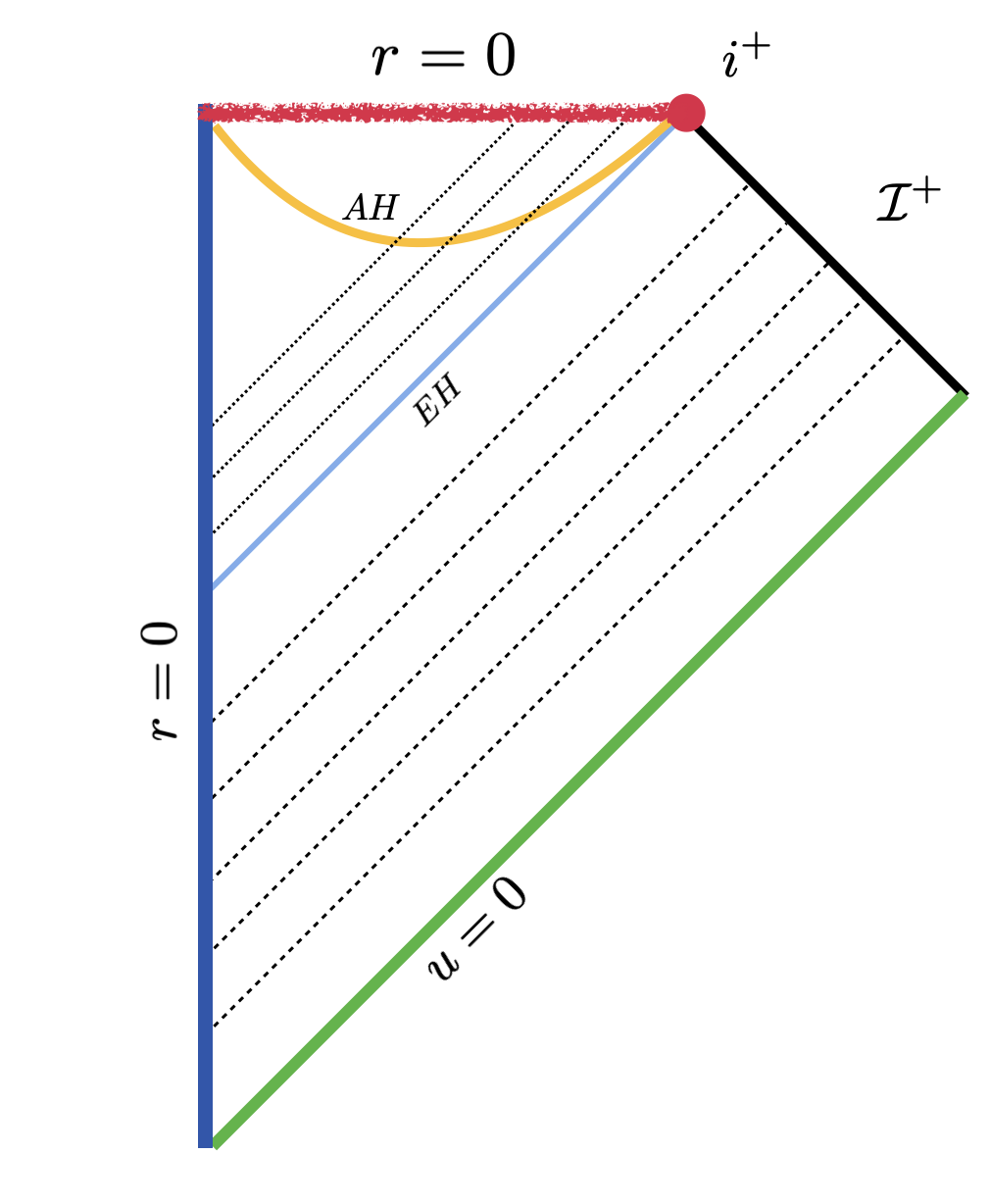}
\caption{\label{conf_supercritical} Conformal diagram of a supercritical evolution. 
The outgoing null hypersurface $u=0$  is the initial data surface extending to null infinity ${\cal I}^+$. $EH$ is the future event horizon $H^+$ which extends to
timelike  infinity $i^+$. The hypersurface $AH$ is the apparent horizon. {The individual outgoing null cones reconverge to point caustics whose locus traces out the final singularity at
$r=0$. }}
\end{figure}

{For supercritical initial data 
our results confirm the well-known results
that the apparent horizon and final singularity
are both spacelike hypersurfaces, as schematically
represented for the compactified spacetime in Fig.~\ref{conf_supercritical} }

We use geometric units in which $c=8\pi G=1$ . Covariant derivatives are denoted by $\nabla$ and partial derivatives are often used in comma notation, i.e. $\partial f/\partial x^a = f_{,a}$. 


 \section{Affine null-metric formalism for the Einstein-scalar field equations in spherical symmetry}
 \label{sec:2}

The  spherically symmetric  metric in affine-null coordinates $u,\lambda,x^A$ is given by \cite{Win2013,TM2019,CW2019,2023PhRvD.107j4010M,2021PhRvD.104h4048G} 
\begin{equation}
ds^2 = -V(u,\lambda)du^2-2dud\lambda +r^2(u,\lambda)q_{AB}(x^C)dx^Adx^B,
\end{equation}
where the outgoing null cones are labeled by the proper time  $u$ along the central
worldline, $\lambda$ is an affine parameter along the spherical congruence of 
outgoing null rays, which are labeled by angular coordinates $x^A$.
Here $q_{AB}$ is the unit sphere metric.

We require regularity along the central wordline, where we use the affine freedom
$\lambda\rightarrow A\lambda +B$
to set $\lambda=0$ and $r_{,\lambda}=1$ on the central worldline.
This gives rise to the local Minkowski coordinate  conditions
\begin{equation}\label{reg_cond}
r(u,0) = 0\;,\;r_{,\lambda}(u,0)=1\;,\; V(u,0)=1\;,\;V_{,\lambda}(u,0)=0.
\end{equation}
(For discussion of these coordinate conditions for the Bondi-Sachs formalism see e.g. \cite{TM2013}).

The Einstein equations for a massless scalar field $\Phi$ are given in terms of
the Ricci tensor $R_{ab}$ by
\begin{equation}
E_{ab}:=R_{ab} -  \Phi_{,a}\Phi_{,b} = 0,
\end{equation}
with components
\begin{subequations}\label{EFE}
\begin{align}
E_{uu}:\;\; &0 =
-\frac{2}{r}r_{,uu}   
+\frac{V}{2r^2}(r^2V_{,\lambda})_{,\lambda} 
+\frac{r_{,\lambda}V_{,u} }{r}
\nonumber\\
  &\qquad
-\frac{r_{,u}V_{,\lambda} }{r}
- \Phi_{,u}^2
\label{E00}\\
E_{u\lambda}:\;\; &0 = \frac{1}{2r^2}(r^2V_{,\lambda})_{,\lambda} 
  - \frac{2r_{,u\lambda}}{r} 
  - \Phi_{,u}\Phi_{,\lambda}\label{E01}\\
E_{\lambda\lambda} :\;\; &0 = -\frac{2}{r}r_{,\lambda\lambda} - (\Phi_{,\lambda})^2  \label{E11} \\
q^{AB}E_{AB} :\;\; &0 
 =  -[V(r^2)_{,\lambda} - 2\lambda- 2(r^2)_{,u}]_{,\lambda}\label{eq:Veqn}
\end{align}
\end{subequations}
and the wave  equation 
$\nabla_a\nabla^a \Phi = 0$, which takes the form
\begin{equation}\label{eq:boxPhi}
0= (r^2\Phi_{,u})_{,\lambda}+(r^2\Phi_{,\lambda})_{,u} -(r^2 V \Phi_{,\lambda})_{,\lambda}\;\;.
\end{equation}
Equations   \eqref{EFE} have a scale symmetry: If $\Phi(u,\lambda)$, $V(u,\lambda)$, $r(u, \lambda)$  is a solution then so is
${\hat \Phi(u,\lambda)=\Phi(\hat u,\hat\lambda)}$, $  
{\hat V(u,\lambda) = V(\hat u,\hat \lambda)}$ and ${\hat r(u,\lambda) =\alpha^{-1}  r(\hat u,\hat \lambda)}$,
where $(\hat u,\hat \lambda)= (\alpha u, \alpha \lambda)$.

If $E_{\lambda\lambda}$, $q^{AB}E_{AB}$ and $\nabla_a\nabla^a \Phi = 0$ hold everywhere, 
then the  Bianchi identities imply that (i) $E_{u\lambda}$ holds trivially and (ii)
\begin{equation} \label{eq:EEuu_bianchi}
 0 = (r^2 E_{uu})_{,\lambda}
\end{equation}
so that if $r^2 E_{uu}$ holds for one value $\lambda$, it holds everywhere. Therefore,
since regularity requires   $r^2 E_{uu}$ to vanish at the central word line
where $r=0$, we need only enforce the three {\bf main equations} $E_{\lambda\lambda}$, $q^{AB}E_{AB}$  and $\nabla_a\nabla^a \Phi = 0$ in a numerical simulation.

The Einstein equation  (\ref{eq:Veqn}) can readily be integrated to find
\begin{equation}
V =\frac{C(u) +2\lambda +2(r^2)_{,u}}{(r^2)_{,\lambda}},
\end{equation} 
where the coordinate conditions (\ref{reg_cond}) require that the  constant $C=0)$.
Consequently,
\begin{equation}\label{Vsol}
V =\frac{\lambda }{rr_{,\lambda}}
+2\frac{ r_{,u}}{r_{,\lambda}} .
\end{equation}

For future reference, the Ricci scalar, $\mathcal{R} = g^{ab}R_{ab}$ is given by
\begin{equation}\label{RS}
    \begin{split}
        r^2\mathcal{R} = & 4 r_{,\lambda} r_{,u}+ 8 r r_{,u\lambda} - 4 r \left( r_{,\lambda} V \right)_{,\lambda} \\
        & -2 (r_{,\lambda})^2 V-r^2 V_{,\lambda\lambda}+2 ,
    \end{split}
\end{equation}
which according to (\ref{Vsol})  reduces to
\begin{equation}\label{eq:Ricci_simple}
    \mathcal{R} = \frac{2\lambda r_{,\lambda}}{r^3} - V_{,\lambda\lambda} - \frac{2}{r^2}.
\end{equation}

Using  \eqref{Vsol},
the line element for the affine null metric with regular origin becomes
\begin{equation}\label{ds2_vsol}
    ds^2 = -\Big(\frac{\lambda+2rr_{,u}}{rr_{,\lambda}}\Big)du^2
    -2d\lambda du
    +r^2q_{AB}dx^Adx^B,
\end{equation}
which shows that the metric is entirely determined once $r$ is known. 
Given regular initial data $r=r(u_0,\lambda)$ on the null hypersurface $u=u_0$, integration of \eqref{E11} then  determines the initial value of the scalar field according to
\begin{equation}
    \Phi(u_0,\lambda) = \int_{0}^\lambda 
    d\tilde \lambda
    \sqrt{-\frac{2r_{,\tilde\lambda\tilde\lambda}(u_0,\tilde \lambda)}{r(u_0,\tilde \lambda)}} \;\;.
\end{equation}
We use this procedure for construction of initial data in Sec.~\ref{inidata}. 


\subsection{Some general physical properties}\label{sec:gen_prop}

In spherical symmetry the Misner-Sharp mass 
\begin{equation}\label{def_M_ms}
    m = \frac{r}{2} (1- g^{ab}r_{,a}r_{,b}) = \frac{1}{2}(r -\lambda r_{,\lambda}) ,
\end{equation}
{where the second equality results from using (\ref{Vsol}) to
replace $r_{,u}$. } This 
provides an invariant quasi-local definition of mass, which is related to the Bondi mass $M_B$
of an asymptotically flat null cone by 
\begin{equation}
M_B = \lim_{\lambda\rightarrow\infty}m.    
\end{equation}
Asymptotic flatness also implies 
\begin{align}\label{phi_asympt}
    \Phi(u,\lambda) = \frac{\Phi_{[1]}(u)}{\lambda} + \frac{\Phi_{[2]}(u)}{\lambda^2} + \mathcal{O}(\lambda^{3})
\end{align}
where the gauge freedom $\Phi\rightarrow \Phi+const$ is used to set $\Phi(u,\lambda) \rightarrow 0$ as $\lambda\rightarrow\infty$.
Then  \eqref{E11} implies
\begin{equation}\label{r_asympt}
    r(u,\lambda) = H(u)\lambda + 2M_B(u) + \mathcal{O}(\lambda^{-1})
\end{equation}
where $H=\lim_{\lambda\rightarrow\infty}r_{,\lambda}$ is a function of integration.
The Bondi time $u_B$ for an inertial observer at infinity is related to the proper time $u$ at the origin by
\begin{equation}
  u_{B,u} = \frac{1}{H}\;.
\end{equation}
For  $H<1$, the Bondi time for an observer at infinity runs faster and it takes an infinite Bondi time to reach the
event horizon but only a finite central time. Consequently, an infinite  observer  is red shifted with respect to  a freely falling observer at the origin. An event horizon forms
at a time $u_E$ when
\begin{equation}\label{event_horizon}
   \lim_{u\rightarrow u_E}H=0,
\end{equation}
 i.e. when the redshift becomes infinite and a black hole forms.
 Thus $H$ monitors the communication between an observer
 at the central world and an inertial observer at null infinity.
When $H=0$ such communication is not possible and the central world line enters an event horizon.
At that time  \eqref{r_asympt} shows that
$\lim_{u\rightarrow u_E,\lambda\rightarrow \infty}r_{,\lambda}=0$.
Here $r_{,\lambda}$ is related to the expansion $ \Theta^+ $ of the outgoing null
cones by
\begin{equation}
    \Theta^+ = \frac{2r_{,\lambda}}{r}.
\end{equation} 
   
Upon further evolution, when the central worldline enters the black hole, the null
cones form an apparent horizon where
$r_{,\lambda}(u,\lambda_A)=0$ and the expansion vanishes at a finite affine value $\lambda_A$. 
Indeed,  \eqref{E11} shows that $r_{,\lambda}$ is a monotonically decreasing function of $\lambda$ so that the expansion becomes negative  and the $\lambda_A(u)$ 2-spheres are  trapped
for $\lambda>\lambda_A$. {The individual outgoing null cones then
reconverge to a point caustic at some finite affine value
$\lambda_C>\lambda_A$, where
$ \Theta^+ \rightarrow -\infty$. The final singularity traced
out by the locus of these caustics  is spacelike.}

According to \eqref{def_M_ms}, the mass of the apparent horizon is given by ${m(u)=r(u, \lambda_A)/2}$.
Also,  the regularity  of the metric  component $g_{uu}$ at the apparent horizon implies, via \eqref{Vsol}, that 
\begin{equation}
\label{ }
\lambda_A(u) = -2r(u,\lambda_A)r_{,u}(u,\lambda_A) .
\end{equation}  
The normal vector to the apparent horizon
$r_{,\lambda}(u,\lambda_A)=0$  has norm
\begin{equation}
g^{ab}\nabla_a r_{,\lambda}\nabla_b  r_{,\lambda}  |_{\lambda=\lambda_A}
= r_{,\lambda\lambda}(Vr_{,\lambda\lambda} 
-2r_{,\lambda u} )  |_{\lambda=\lambda_A} .
\end{equation}
But  \eqref{Vsol} implies 
\begin{equation}
rr_{,\lambda} V=\lambda+2r r_{, u} 
\end{equation}
so that
\begin{equation}
  rr_{,\lambda\lambda} V \bigg|_{\lambda_A} =1+2rr_{,\lambda u} \bigg|_{\lambda_A}.
  \label{rddrV_deriv}
\end{equation}
Thus the norm is given by
\begin{equation}
g^{ab}\nabla_a r_{,\lambda}\nabla_b  r_{,\lambda}  \bigg|_{\lambda_A}
= \frac{r_{,\lambda\lambda}}{r} \bigg|_{\lambda_A} = -\frac{\Phi_{,\lambda}^ 2}{2}\bigg|_{\lambda_A} <0
\end{equation}
so that the apparent horizon is a spacelike hypersurface, as expected from the general theory.

The numerical algorithm is designed to evolve in a timelike direction, which requires that the metric variable $V>0$.  Thus the sign of  $V$ is
important. In Sec.~\ref{sec:reg}
we show that $V>0$ in the region where $r>\lambda$ (see (\ref{eq:vo})). Near the origin, $r\approx \lambda$
and $r_{,\lambda }\approx 1$. Furthermore,
throughout a region containing
the apparent horizon, where $r$ attains its maximum value, $r$  is a monotonically increasing function
of $\lambda$. Thus $r>\lambda$ 
and  $V >0$ throughout the exterior asymptotically flat region and
in a region inside the event horizon
 $0\le \lambda < \lambda_A <\lambda_V$, which
contains the apparent horizon.

 At the apparent horizon, \eqref{rddrV_deriv} implies
\begin{equation}
  1+2rr_{,\lambda u} \bigg|_{\lambda_A}<0.
\end{equation}
The value $\lambda_V$ where $V=0$ satisfies
\begin{equation}
(\lambda+2r r_{, u})|_{\lambda_V} =0\;\;.
\end{equation}
Thus  $\lambda+2r r_{, u}$ must have
a maximum between $\lambda_A$ and $\lambda_V$.
On the assumption that the gradient $r_{,a}$ remains finite at the final singularity
as $r\rightarrow 0$,  \eqref{Vsol} implies
$V \rightarrow \lambda /(rr_{,\lambda})\rightarrow -\infty$ as $\lambda\rightarrow \lambda_C$,
so that the $u$-direction eventually becomes spacelike. 
This issue is discussed further in Sec.~\ref{sec:prop_idata}.

The norm of the $r=const$ hypersurfaces  is given by
\begin{equation}
 g^{ab} \nabla_a r\nabla_br = V r_{,\lambda}r_{,\lambda} -2 r_{,\lambda}r_{,u}
      =  \frac{\lambda r_{,\lambda}}{2r} .
\end{equation}
Since $ r_{,\lambda}$ is a monotonically decreasing function of $\lambda$
these hypersurfaces are timelike inside the apparent horizon, null at the apparent
horizon and spacelike past the apparent horizon. In particular, the final singularity
at $\lambda=\lambda_C$ is spacelike.


\subsection{Main equations as a hierarchy}

Because of the appearance of $r_{,u}$ in \eqref{eq:Veqn}, the 
main equations \eqref{E11}, \eqref{eq:Veqn} and \eqref{eq:boxPhi} do not have
the hierarchical structure of the Bondi-Sachs equations which can be integrated in
sequential order \cite{BMS,TW1966,Maedler:2016}.  A method to restore a hierarchy
was  presented  in \cite{CW2019}.
Introduction of the term
\begin{equation}\label{eq:K_def}
K :=  2r_{,\lambda}\Phi_{,u} - 2r_{,u}\Phi_{,\lambda}
\end{equation}
allows the wave equation \eqref{eq:boxPhi}  to be written as the null hypersurface equation
\begin{equation}\label{eq:K_derived}
r\Big(\frac{rK }{r_{,\lambda}}\Big)_{,\lambda} =
 \left(\frac{r \lambda \Phi_{,\lambda}}{r_{,\lambda}}\right)_{,\lambda}\;.
\end{equation}
The $u$-derivative of $\Phi$ is then determined  from $K$,
\begin{equation}
\label{Phiu}
\Phi_{,u} = \frac{K}{2r_{,\lambda}} +\frac{ r_{,u}}{r_{,\lambda}} \Phi_{,\lambda} \;\;.
\end{equation}
Next, the hypersurface equation for $r_{,u}$ results from the $u$-derivative of \eqref{E11}.
This leads to  three  ordinary differential  equations for $r, \; K $ and $r_{,u}$,
\begin{subequations}\label{hier1}
\begin{align}
r_{,\lambda\lambda} = & 
-\frac{r}{2}\Phi_{,\lambda}^2\label{r_lambdalambda_orig}
\\
\Big(\frac{rK}{r_{,\lambda}}\Big)_{,\lambda}=& 
\frac{1}{r}\Big(\frac{r\lambda\Phi_{,\lambda}}{r_{,\lambda}}\Big)_{,\lambda}
\label{Keqn}\\
\Big(\frac{r_{,u}}{r_{,\lambda}}\Big)_{,\lambda\lambda}=&-\frac{r\Phi_{,\lambda}}{2 r_{,\lambda}}\Big(\frac{K}{r_{,\lambda}}\Big)_{,\lambda} ,
\label{r_u_eqn}
\end{align}
\end{subequations}
which can be integrated in sequential order, with $\Phi_{,u}$ then determined from
(\ref{Phiu}).

\subsection{Regularized version of the equations}
\label{sec:reg}

In the exterior asymptotically flat region, where $r_{,\lambda} \ne 0$,  \eqref{hier1} and \eqref{Phiu}
provides a well defined evolution algorithm.
However, for evolution inside a black hole, a regularization  scheme
is necessary to remove singular $1/r_{,\lambda}$  terms in  \eqref{hier1}.
Such terms are not due to coordinate
singularities but seem to be an artifact of the affine-null equations. 
Here we present a brief account of the  regularization procedure in \cite{CW2019}.

Introduction of the variable
\begin{align}
    L =& \frac{rK -\lambda\Phi_{,\lambda}}{r_{,\lambda}} = 2r\Phi_{,u} - rV\Phi_{,\lambda}
\end{align}
allows  us replace  \eqref{Keqn} by
\begin{equation}\label{Leqn}
    L_{,\lambda} = \frac{\lambda}{r}\Phi_{,\lambda} \, ,
  \end{equation}
which can be integrated from $\lambda=0$ using the regularity  condition $L(u,0)=0$ at the origin. 
Here the  Ricci scalar \eqref{eq:Ricci_simple} has a simple expression in terms of the new variable,
\begin{equation}
        \mathcal{R}        = -\frac{L\Phi_{,\lambda}}{r} .
\end{equation}

For the regularisation of  \eqref{r_u_eqn}, we introduce the variable
\begin{equation}\label{def_Q}
Q=\frac{V-1}{\lambda}=  \frac{2rr_{,u} +\lambda-rr_{,\lambda}}{\lambda rr_{,\lambda}} \, ,
\end{equation}
which satisfies 
\begin{equation}
Q_{,\lambda}=  \frac{1}{\lambda^2}- \frac{1}{r^2} + \frac{1}{2}\left(\frac{L}{\lambda} \right)^2 .
\label{Qhypersurface}
\end{equation}
This equation can be integrated from $\lambda=0$ using the regularity condition $Q(u,0)=0$. Note that the right hand side of (\ref{Qhypersurface}) contains the terms
$1/\lambda^2$ and $1/r^2$ which are singular at the origin but combine to form
a regular function. This is handled by  numerical  techniques in Sec.~\ref{sec:num_detail}.

In  summary, the hierarchical evolution system \eqref{hier1} takes the regularized form
\begin{subequations}\label{hier_alt}
\begin{align}
r_{,\lambda\lambda}=&-\frac{r}{2}\Phi_{,\lambda}^2 \label{hier_alt1}\\
L_{,\lambda}=&\frac{\lambda}{r}\Phi_{,\lambda}\label{hier_alt2}\\
Q_{,\lambda}=&  \frac{1}{\lambda^2}- \frac{1}{r^2} + \frac{1}{2}\left(\frac{L}{\lambda}\right)^2
\label{hier_alt3}\\
\Phi_{,u} =&\frac{1+\lambda Q}{2}\Phi_{,\lambda} + \frac{L}{2r} .\label{hier_alt4}
\end{align}
\end{subequations}
The right hand sides of these equations remain regular up to the formation of the
final singularity at $\lambda=\lambda_C$, where $r=0$. However, numerical
treatmen of the right hand  side
of  (\ref{hier_alt3}) requires special attention near the origin, where cancellations lead to
\begin{equation}
Q_{,\lambda}(u, \lambda)=\frac{2}{3}\Phi_{,\lambda}^2(u, 0) +\mathcal{O}(\lambda) \; .
\end{equation}

Given initial data $\Phi(u_0, \lambda)$ on a null cone $u_0$, a numerical
evolution scheme proceeds  by integrating  (\ref{hier_alt1}) - (\ref{hier_alt3})
sequentially. Then  \eqref{hier_alt4} provides a finite difference
approximation to update $\Phi(u_0+\Delta u, \lambda)$. This procedure is then be iterated
into the future.

In an exterior asymptotically flat region, the hypersurface integrations proceed from
$\lambda=0$ to $\lambda=\infty$. However, inside a black hole, the final singularity
is formed at a finite value  $\lambda=\lambda_C$.
We can rewrite \eqref{hier_alt4} as the transport equation
\begin{equation}
 \Phi_{,u} -\frac{V}{2}\Phi_{,\lambda} =  \frac{L}{2r}\;\;.
\end{equation}
Consequently, for a timelike outer boundary with $V>0$,
the transport would be in the inward $\lambda$ direction so that an outer boundary condition
would be necessary. In the exterior, the compactified outer boundary at infinity is null so that no boundary condition is necessary.

In the black hole interior, in order to avoid introducing spurious
outer boundary data we set the outer boundary at the apparent horizon, $\lambda=\lambda_A$,
which is spacelike so that no outer boundary condition is needed. In order to see that $V>0$ inside the apparent horizon so that the evolution
proceeds in a timelike direction we refer back
to Sec.~\ref{sec:gen_prop} to note  $r>\lambda$ for $0\le \lambda\le \lambda_V$, where
$V(u,\lambda_V)=0$. Thus
\eqref{Qhypersurface} implies $Q_{,\lambda}\ge 0$
in that region and, since
$Q(u,0)=0$, it follows that $Q\ge 0$.
Thus, according to
\eqref{def_Q}, 
\begin{equation}
V=1+\lambda Q>0\, ,\quad 0< \lambda_A<\lambda_V 
\label{eq:vo}
\end{equation}
so that $V>0$ inside the apparent horizon.

\section{Initial data}\label{inidata}

We recall from Sec.~\ref{sec:2} that  all metric functions can be determined 
from $r$. We introduce the  initial data
\begin{equation}
\label{rAEI}
r(u=0,\lambda) = \lambda -\frac{b^2\lambda^3}{(a+ \lambda)^2}
=  \lambda\left[\frac{(a+\lambda)^ 2-b^2\lambda^2}{(a+\lambda)^ 2}\right]
\end{equation}
depending on two positive parameters $a$ and $b$ 
for which all the hypersurface equations can be
integrated to determine $V$, $\Phi$ and $\Phi_{,u}$ analytically.
This data satisfies the local Minkowski conditions \eqref{reg_cond}  for $r$ at the origin.

From the corresponding derivative,
\begin{equation}
r_{,\lambda}(0,\lambda; a, b) =\frac{(1-b^2 )(\lambda^3 + 3a \lambda^2) + 3a^2\lambda + a^3}{(a+ \lambda)^3}\, ,
\end{equation}
the hypersurface equation \eqref{r_lambdalambda_orig} gives the derivative of the
initial scalar field
\begin{equation}
    \Phi_{,\lambda}(0,\lambda) =  \frac{ ab\sqrt{12} }
      {(a+\lambda)[(a+\lambda)^2-b^2 \lambda^2]^{1/2}} \, ,
\end{equation}
whose integral determines the initial scalar field 
 \begin{align}\label{Phi_data}
    \Phi(0,\lambda) =&
    \sqrt{12} \,  \left[ \sin^{-1} \bigg(  
      \frac{b \lambda}{a+\lambda} \bigg )
      - \sin^{-1}b \right] .
\end{align}
(Here we have chosen the integration constant such that $\lim_{\lambda=\infty}\Phi=0$
for asymptotically flat initial data. For initial data on a  null hypersurface inside a black hole,
where $b>1$, we use the gauge freedom to drop the $\sin^{-1}b$ term so that
$ \Phi(0,\lambda)$  remains a well-defined real function.)

The $L$, $Q$ and $\Phi_{,u}$ hypersurface equations together with \eqref{def_Q} give
\begin{widetext}
\begin{subequations}\label{ana_idata}
\begin{align}
L(0, \lambda) =& \frac{\sqrt{12} b\lambda}{\sqrt{(a+\lambda)^2-b^2\lambda^2}} \label{iData_L}\\
V(0, \lambda) =& 
   1+ \lambda\left[ \frac{9b}{4a}  \ln \bigg (
  \frac{1+(1+b)\frac{\lambda}{a}}
  {1+(1-b)\frac{\lambda}{a}}
  \bigg) 
  -\frac{1}{2}\frac{b^2}{a^2}\frac{\lambda(1+\frac{\lambda}{a})} 
  {[(1+\frac{\lambda}{a})^2-b^2 \left(\frac{\lambda}{a}\right)^2]}\right]\;\;
 \label{iData_V}\\
\Phi_{,u}(0,\lambda) =& \frac{b\sqrt{3}}{
       \sqrt{(a+\lambda)^2-b^2\lambda^2) }}
       \left \{ \frac{9b\lambda}{4(a+\lambda)} 
       \ln\left[  \frac{a+(1+b)  \lambda}
      {a+(1-b) \lambda} \right ]
       +\frac{1}{2} +\frac{a}{a+\lambda}
    +\frac{(a+\lambda)^2}{2[(a+\lambda)^2 -b^2\lambda^2]} \right\}
     \label{iData_dphidu}
\end{align}  
which also determine the Ricci scalar $\mathcal{R}$ and  $r_{,u}$.
\begin{align}
\mathcal{R}(0, \lambda) = &
-\frac{12b^2(1+\frac{\lambda}{a})}{[(1+\frac{\lambda}{a})^2-b^2\left(\frac{\lambda}{a}\right)^2]^2}\\
r_{,u}(0, \lambda) =& 
- \frac{9b\lambda r_{,\lambda} }{8a} \ln \bigg (
  \frac{a+(1+b)\lambda}
  {a+(1-b)\lambda}
  \bigg) 
  -\frac{b^2\lambda^2(3a+\lambda)^2} 
  {4a(a+\lambda)^3}  .
\end{align}
\end{subequations}
\end{widetext}

These explicit values of the fields facilitate
measuring the convergence and accuracy of the numerical
integrators for the hypersurface equations. 


\subsection{Properties of the initial data}\label{sec:prop_idata}

The initial data have the general scaling behavior discussed in Sec.~\ref{sec:2}, e.g.
$r(\alpha \lambda; a, b) = \alpha r\left(\lambda;  \frac{a}{\alpha},b\right)$
and similarly  ${V(\alpha\lambda;a,b) = V(\lambda;\frac{a}{\alpha}, b)}$.
In principle, one can set $a=1$ without loss of generality, but other choices
are beneficial for numerical purposes, as seen in Sec.~\ref{sec:num_detail}.

The choice of $b$ determines the nature of the initial null hypersurface:
\begin{itemize}

\item $b=0$ determines a flat space null cone.

\item  $b <1$ determines an asymptotically flat  null hypersurface 

\item $b=1$ determines an event horizon

\item  $b>1$ determines a null  hypersurface inside a black hole.

\end{itemize}

This nature can be inferred from calculating the asymptotic expansion \eqref{r_asympt} of \eqref{rAEI} which gives
\begin{equation}
H=1-b^2\;\;,\;\;M_B = 2ab^2.
\end{equation}

The three roots of the cubic equation obtained by setting $r=0$ in \eqref{rAEI}, are
\begin{equation}
    \lambda_0 = 0\;\;,\;\;
    \lambda_1 = -\frac{a}{b+1}\;\;,\;
    \lambda_2 = \frac{a}{b-1}\;\;,\; 
\end{equation}
Here $\lambda_0$ is the known caustic at the vertex of the nullcone. 
Next, $\lambda_1 $ is unphysical, i.e. it is negative so outside the physical domain. The third root $\lambda_2$ is ony physical for $b>1$ and represents the caustic
$$\lambda_C=\frac{a}{b-1}\;\;,\
$$
at the final singularity of the black hole. From \eqref{def_M_ms},  the Misner-Sharp mass of the singular caustic is ${m( \lambda_C) = a/b}$.

\begin{figure}
\includegraphics[scale=0.4]{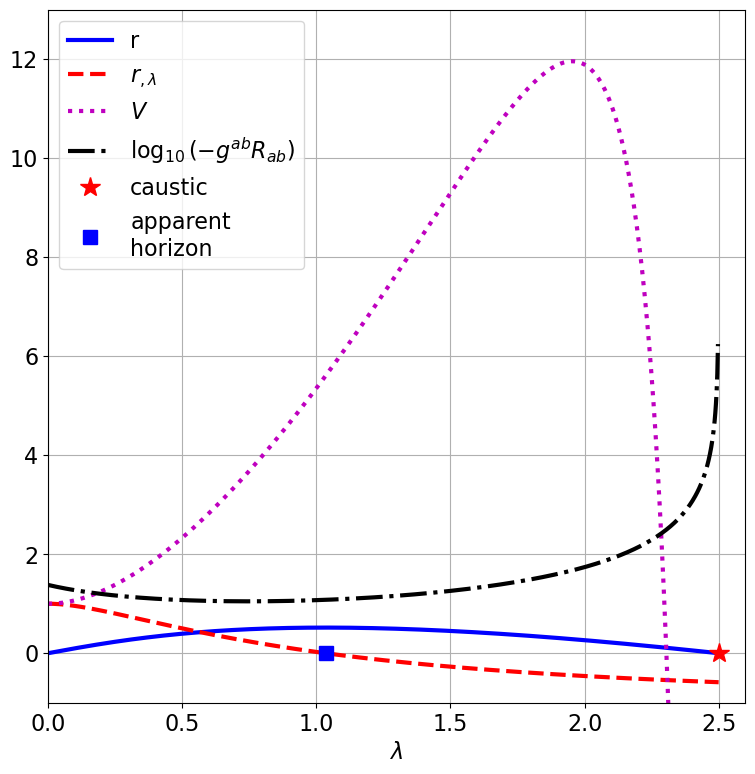}
\caption{\label{iData}Inside event horizon initial data for $a=2.5(\sqrt{2}-1)$ and $b=\sqrt{2}$. The singular caustic is at $\lambda_C=2.5$ and the apparent horizon is at $\lambda_A=2.5(\sqrt{2}-1)\approx1.04$.}
\end{figure}

The local behavior off  the Ricci scalar $\mathcal {R}$ and $V$  near the vertex $\lambda_0$
is given by the expansion
\begin{align}
\mathcal{R}(0, \lambda) &= \frac{-12b^2}{a^2}+\mathcal{O}(\lambda-\lambda_0)\\
V(0, \lambda)&=
1+\mathcal{O}[(\lambda-\lambda_0)^2]
\end{align}
and near the caustic  $\lambda_C$ by the expansion
\begin{align}
\mathcal{R}(0, \lambda) &
= 
-\frac{3b}{b-1}\left(\lambda -\lambda_C\right)^{-2}
+\mathcal{O}\left[\left(\lambda -\lambda_C\right)^{-1}\right]\label{Ric_near_caustic}\\
V(0, \lambda)&=
 - \frac{ab^2}{4(b-1)^3}\left(\lambda - \lambda_C\right)^{-1}
 +\mathcal{O}(1)  .
\end{align}
The behavior near  $\lambda_0$  reflects the regularity of the vertex.
As anticipated by the discussion in Sec.~\ref{sec:gen_prop},  $V\rightarrow -\infty$ at $\lambda_C$ 
and the Ricci scalar is singular. At the apparent horizon $V(\lambda_A) >0$ and the
$u$-direction is time like. At a value $\lambda_V >\lambda_A$, $V$ changes sign
and the $u$-direction becomes spacelike. 
See Fig.~\ref{iData}, which  displays the 
behavior of $V(u,\lambda)$ on null hypersurfaces inside an event horizon. Fig.~\ref{iData} also shows that $r_{,\lambda}$ is finite at the caustic, while the Ricci scalar $g^{ab}R_{ab}$ and $V$ go to negative infinity. 


\subsection{Location of the apparent horizon}

For the black hole data, between the two caustics at 
$\lambda=0$ and $\lambda_C$, $r$ attains a maximum at the apparent horizon where $r_{,\lambda}(\lambda_A)=0$.
The location of the  apparent horizon is found by setting
$r_{,\lambda}(\lambda_{A})=0$, which leads to the cubic equation 
\begin{equation}
0=(1-b^2)(\lambda^3_{A} + 3a\lambda^2_{A})+3a^2\lambda_{A}+a^3 .
\end{equation}

Setting $\lambda_A=a/(y_A-1)$, the cubic takes the reduced form
\begin{equation}\label{eq:cubic}
      P(y)=y^3-3b^2y +2b^2 =0.
\end{equation}

The three roots are given by
\begin{equation}
\label{ }
y_k=2b \cos\left(\frac{\psi}{3}+\frac{2\pi k}{3}\right), \quad k=0,1,2
\end{equation}
where $\cos\psi= -1/b$. 
A physical root must satisfy $0<\lambda_A<\lambda_C = a/(b-1)$ or
$y_A>b>1$.
Since  $P(0)=2b^2>0$ and   $P(y)$ has a  maximum at $y=-b$
and  a minimum at $y=+b$, where
$P(+b)=-2b^2( b-1)<0$, there is only one real root  with $y>b>1$. 
This corresponds to $y_{k=0}$ so that 
\begin{equation}
\label{ }
y_A =2b \cos(\psi/3) =2b \cos\left[\frac{1}{3}\arccos\left(-\frac{1}{b}\right)\right] \;.
\end{equation}

The cubic has a  particularly simple solution
for $b^2=2$, where  $P(y)=(y-2)(y^2+2y-2)$. 
The relevant root is $y_A=2$ and $\lambda_A=a$.
For this case,
\begin{equation}
\label{ }
V(\lambda_A)=\frac{9\sqrt{2}}{4}\ln \bigg(\frac{2+\sqrt{2})}{2-\sqrt{2})}\bigg) >0
\end{equation}
and  
\begin{equation}
\label{ }
r_{,u}(\lambda_A)=-1.
\end{equation}

\section{From the physical field equations to their representation on a computer}
\subsection{A generalized spatial grid}

For the numerical simulations, we represent  the affine parameter by a  grid coordinate $x\in[-1,1]$ so that $\lambda = \lambda(x)$, while requiring
$\lambda(-1) = 0$ and $\lambda^\prime(-1)\neq0$. 
With respect to the new grid variable \eqref{hier_alt} becomes
\begin{subequations}\label{reg_hier_gen_grid}
\begin{eqnarray}
    0&=& r_{,xx} - (\ln |\lambda^{\prime}|)^\prime  r_{,x} +\frac{r}{2}\Phi_{,x}^2\\
    L_{,x}&=&\frac{\lambda}{r}\Phi_{,x}\\
 \frac{ Q_{,x}}{\lambda^\prime}
 &=&
 \frac{1}{\lambda^2} - \frac{1}{r^2}+ \frac{1}{2}\left(\frac{L}{\lambda}\right)^2\\
  \Phi_{,u} & = &\frac{1+\lambda Q}{2\lambda^\prime}\Phi_{,x}+ \frac{L}{2r} .
 \end{eqnarray}
\end{subequations}
Two grid functions $\lambda(x)$ are employed: (i) a linear grid function covering the finite
$\lambda$  domain inside a black hole: and  (ii) a compactified grid function covering the infinite
$\lambda$ domain on an asymptotically flat  null hypersurface.

\subsubsection{Linear grid function  - interior region}
The linear grid function 
\begin{equation}\label{lambda_lin}
\lambda_I(x) = A(x+1),
\end{equation} 
with $A>0$, allows us to determine the fields for ${\lambda\in[0, 2A]}$, The hierarchy \eqref{reg_hier_gen_grid} then becomes
\begin{subequations}\label{reg_hier_linear}
\begin{eqnarray}
    0&=& r_{,xx}   +\frac{r}{2}\Phi_{,x}^2\\
    L_{,x}&=&\frac{A(x+1)}{r}\Phi_{,x} \label{eq:L_num_lin}\\
    Q_{,x}
 &=&
 \frac{1}{A(1+x)^2} - \frac{A}{r^2}+ \frac{1}{2A}\left[\frac{L}{(1+x)}\right]^2 \label{hyp_Q_lin}\\
\Phi_{,u} & = &\frac{1}{2A}\left[1+A(1+x) Q\right]\Phi_{,x}+ \frac{L}{2r}.
\label{eq:dphidu_num_lin}
\end{eqnarray}
\end{subequations}
The inner boundary conditions for the fields are
\begin{equation}
\begin{split}
&r(-1) = 0\;\;,\;\;r_{,x}(-1)=A\;\;,\;\;L(-1)=Q(-1) = 0\;\;,\\
&\Phi_{,u}(-1) =\frac{1}{A}\Phi_{,x}(-1).
\end{split}
\end{equation}
In terms of the $x$ coordinate, ${r_{,\lambda} = r_{,x}/A}$ and 
$$r_{,u}  = \frac{1}{2}\left(\frac{A(1+x)}{r}+\frac{r_{,x}}{A}[1+A(1+x)Q]\right) .
$$
In order to start up the integration at $x=-1$ and enforce regularity at the origin,
we numerically determine  the derivative $\Phi_{(1)}(u) = \Phi_{,x}(u, -1)$ of the data at the origin.
Then the right hand sides of \eqref{eq:L_num_lin} - \eqref{eq:dphidu_num_lin}
have the boundary values
\begin{align}
L_{,x}|_{x=-1}&= \Phi_{(1)}(u)\\
Q_{,x}|_{x=-1}&= \frac{1}{3}\Phi_{(1)}^2(u)\\
\Phi_{,u}|_{x=-1}&= \frac{1}{A}\Phi_{(1)}(u).
\end{align}


\subsubsection{Compactified grid function  - exterior region}

For the compactified grid function, we set
\begin{equation}\label{lambda_compact}
    \lambda_{II}(x):=\lambda(x) = 2A\frac{1+x}{1-x} ,
\end{equation}
with $\lambda^\prime = 4A(1-x)^{-2}$,
which  maps  the infinite $\lambda$ domain into $(-1\le x\le+1)$ 

In order to regularize terms in \eqref{reg_hier_gen_grid} of the form $1/(1-x)^k$ 
which are singular at $x=1$ we introduce the auxiliary variable
\begin{equation}
    R=  (1-x)r ,
\end{equation}
with boundary values
\begin{align}
r(u, \lambda=0)= 0\;\;\; \rightarrow&\;\;\;R(u, x=-1)= 0\\
r_{,\lambda}(u, \lambda=0)= 1\;\;\; \rightarrow&\;\;\;R_{,x}(u, x=-1)= 2A\;.
\end{align} 
The derivatives of $r$ then transform into 
\begin{align}
    r_{,\lambda} = & \frac{(1-x)R_{,x}+R}{4A}
        \\
    (1-x)r_{,u}=& \frac{1}{2}\left\{
    \frac{1}{2A}[(1-x)R_{,x}+R] - \frac{A(1-x^2)}{R}
    \right\}  .
\end{align}

With these transformations,  \eqref{reg_hier_gen_grid}  becomes
\begin{subequations}\label{reg_hier_compact}
\begin{eqnarray}
0&=&R_{,xx} +\frac{\Phi_{,x}^2}{2} R\\
L_{,x}&=& \frac{2A(1+x) }{R}\Phi_{,x}\label{eq:L_num_comp}\\
 Q_{,x}
 &=&
 \frac{1}{A(1+x)^2} - \frac{4A}{R^2}+ \frac{1}{2A}\left[\frac{L}{(1+x)}\right]^2
 \label{hyp_Q_compact}\\
\Phi_{,u} 
&=&
\frac{(1-x)}{2}\left\{\left[1-x+2A(1+x)Q\right]\frac{\Phi_{,x}}{4A}+\frac{L}{R}\right\} .
\nonumber
\\
\label{eq:dphidu_num_comp}
\end{eqnarray}
\end{subequations}
The right hand sides of (\ref{eq:L_num_comp}) -  (\ref{eq:dphidu_num_comp})
have finite values at $x=-1$,
\begin{align}
L_{,x}|_{x=-1}&= \Phi_{(1)}(u)\\
Q_{,x}|_{x=-1}&= \frac{4}{3}\Phi_{(1)}^2(u)\\
\Phi_{,u}|_{x=-1}&= \frac{\Phi_{(1)}(u)}{A} .
\end{align}


\section{Some details on the numerical implementation}\label{sec:num_detail}

We solve the set of hypersurface-evolution equations using a pseudo-spectral method coupled with either an explicit second or third order scheme for the time evolution. 
We discretize the interval $-1\le x\le+1$ into  $N+1$ Gauss-Lobatto points 
\begin{equation}
    x_i = -\cos\left(\frac{i\pi }{N}\right)\;\;,\;\;i=0,1,..., N.
\end{equation}

The central time at the origin is discretized according to 
\begin{equation}
u_n = u_{init} + n \Delta u,
\end{equation}
where $u_{init}$ is the initial time and the time step $\Delta u$ is given by  
\begin{equation}
\label{ }
\Delta u = \frac{u_{fin}-u_{init}}{(N-1)^2},
\end{equation}
where $u_{fin}$ is the final evolution time.
{For spectral methods, note that stability requires the time
step be $ O(N^{-2} )$ \cite{Hesthaven}, as opposed to
$O(N^{-1})$ for finite difference methods.\footnote{{The collocation points are closest spaced near the edge of
the grid, i.e.
 $x_1-x_0 = -\cos(\pi/N)+1 \propto N^{-2}$.}}  }
The scalar field $\Phi$ as well as
\begin{equation} \label{f_def}
f\in\{r, L, Q, \Phi_{,u}\}
\end{equation} 
at a given time $u_n$  are represented
by $N+1$ discrete values on the collocation points $x_i$,
\begin{equation}
\label{ }
f^n_i:= f(u_n, x_i) = \sum_{k=0}^N f_k(u_n)T_k(x_i)
\end{equation}
where $T_i(x)$ are Chebysheff polynomials of the first kind.  

In comparison to previous affine-null implementations  \cite{CW2019} we use
only one spatial  domain of Gauss-Lobatto points and functions are expanded in Chebysheff polynomials of the first kind. Ref. \cite{CW2019} used a two domain spectral decomposition of the $\lambda$ axis and expanded  fields in rational Chebysheff polynomials. Our approach uses half the number of Fast Fourier Transforms (FFTs)  and  matrix multiplications  needed  for the mapping between the Chebysheff coefficients and the functions evaluated at the collocation points. Another variation from \cite{CW2019} is
the compactifiation of null infinity, as implemented in other characteristic codes  \cite{2002PhRvD..65h1501L,Purrer:2004nq}, the Pitt code \cite{2012LRR....15....2W} or  the SpEC code\cite{2015CQGra..32b5008H,2016CQGra..33v5007H}. 

The  code is written in Python, in the framework of the \texttt{anaconda} package \cite{anaconda}. Python performace is improved by employing the \texttt{numba} library \cite{lam2015numba} and the BLAS/LAPACK wrappers of \texttt{scipy} \cite{2020SciPy-NMeth}. 

The implementation of the spectral method follows the review of \cite{Townsend,Olver_Slevinsky_Townsend_2020}.
The initial null data for $\Phi$ are represented on the appropriate grid function
\eqref{lambda_lin} (for the interior of the event horizon) or \eqref{lambda_compact} (for the exterior).
 We then solve the hypersurface equations \eqref{reg_hier_linear} using \eqref{lambda_lin} for the interior or \eqref{reg_hier_compact} for the exterior using \eqref{lambda_compact}.

In order to regularize the right hand sides of \eqref{reg_hier_linear} and  \eqref{reg_hier_compact}, in particular \eqref{hyp_Q_lin} and \eqref{hyp_Q_compact},
near the origin $x=-1$ , we use a {fifth} order Taylor series to fill the values of the Gauss-Lobatto points
inside a word tube $[-1\le x \le -1+\Delta x]$. 
The Taylor series coefficients were determined by successively applying a first order derivative operator to the data  $\Phi(u_n, x_i)$ on the
complete $u=u_n$  null cones.  {{This spectral operator ${\cal D}$ is the
Chebysheff first derivative using a FFT, as described in \cite{Trefethen}. The $p^{th}$ order derivative ${\cal D}^ {(p)}$ of a function $f$ is then given by
\begin{equation}
{\cal D}^ {(p)} f(x_i) = \underbrace{{\cal D}\hdots{\cal D}}_{\mbox{$p$ times}} f(x_i).
\end{equation} }
An optimum value $\Delta x\approx10^{-2}$ was  found by  numerical measurement
of the ${\cal L}_2$ error norms given by \eqref{L2norm}
\begin{equation}\label{L2norm}
\mathcal{L}_2(f) = \sqrt{ \int_{-1}^{1}|f_{num}(x) - f_{ana}|^ 2 dx }
\end{equation}
of  the  fields $f$ defined in \eqref{f_def} between their numerical values, $f_{num} $, and their analytic expressions, $f_{ana}$,  as  given in  \eqref{rAEI}, \eqref{Phi_data}-\eqref{iData_dphidu} with $a=\sqrt{3}/2$ and $b=1/4$. 
Fig.~\ref{L2_norms} displays exponential 
convergence of the ${\cal L}_2$ error norms for the fields $f$.
The error in $\Phi_{,u}$, saturates at round-off error $\approx 2\cdot10^{-14}$.
\begin{figure}[h]
\includegraphics[width=0.45\textwidth]{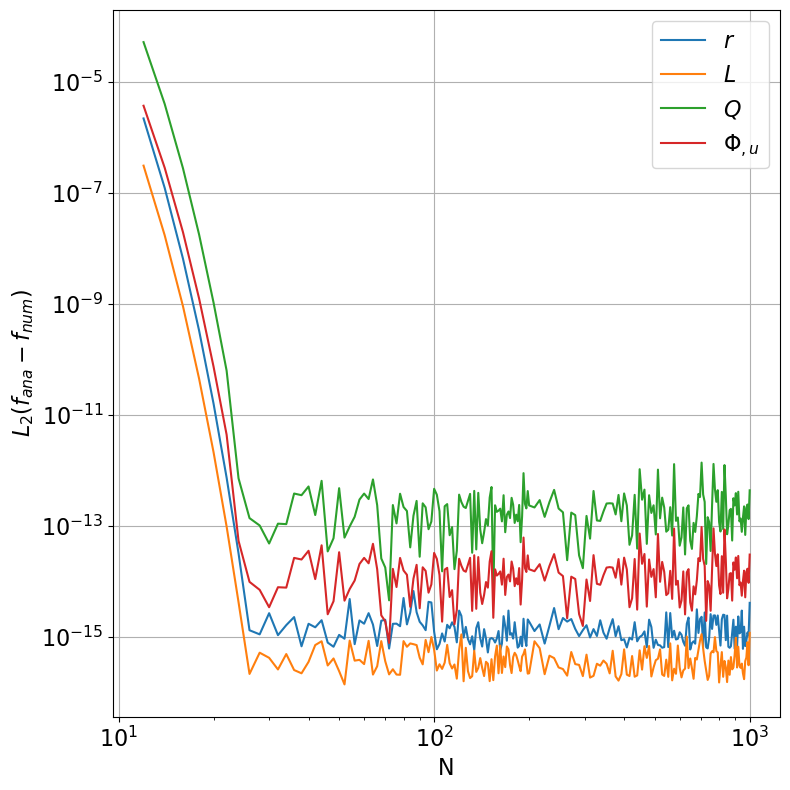}
\caption{\label{L2_norms} $\mathcal{L}_2$ error norms with respect to analytic initial data
for the fields $f$ at different resolutions N. 
} 
 \end{figure}

As a time integrator for the evolution we use either second  or  third order  Runge-Kutta schemes, in particular the (a.k.a Shu-Osher scheme \cite{1988JCoPh..77..439S}) strong stability  conserving Runge-Kutta methods SSPRK22 and SSPRK33 as presented in \cite{Hesthaven}. Aliasing errors arising from high frequency  pseudo spectral modes
are controlled by a classical  2/3 filter. 
{The  convergence  of the time integrators is displayed in Fig.~\ref{figConvTimeLocal} and Fig.~\ref{figConvTimeL2}, for which we chose initial data parameters $(a,b) = (1, 0.94)$. 
The critical time for this initial model is $u^*\approx3$. For these runs we fixed the spatial resolution to 21 collocation points and ran the simulation up until $u=2.975$
while using  the compactified grid function \eqref{lambda_compact}. 
Given the  spatial resolution, stability requires the time step  to be $\Delta u\le  \Delta u_0=1/20^2 $.  
Fig.~\ref{figConvTimeLocal} displays the error in the final 
profiles of $\Phi$ for  simulations with time steps $\Delta u\in{\{2^{-k} \Delta u_0:\; k=0...7 \}}$ using  the second order Runge-Kutta integrator (RK2, upper panel) and  the third order integrator (RK3, lower panel).
The  error profiles are the local error between the final profiles using $\Delta u\in\{ \Delta u_0, \hdots,  du_0/64\}$ and the smallest time step $ \Delta u_0/128$.
The decay rates of the local errors are consistent with second order (a factor $2^2$=4) and third order (a factor $2^3 = 8$). 
This is also seen in the corresponding $\mathcal{L}_2$ norms of those profiles as seen in Fig.~\ref{figConvTimeL2}. 
For reference to the expected behavior of the temporal discretisation error, we plotted the green dotted line  and red dashed line in Fig.~\ref{figConvTimeL2}, which  are proportional to $(\Delta u)^2$ and  $(\Delta u)^3$, respectively. 
We  observe in the  lower panel of  Fig.~\ref{figConvTimeLocal}  and in Fig.~\ref{figConvTimeL2} that the  error for resolution $\Delta u_0/64\approx 4\cdot10^{-5}$  saturates consistent with the maximum error $\lesssim10^{-12}$ of the hypersurface integrator in Fig. ~\ref{L2_norms}.
}
\begin{figure}[h]
\includegraphics[width=0.45\textwidth]{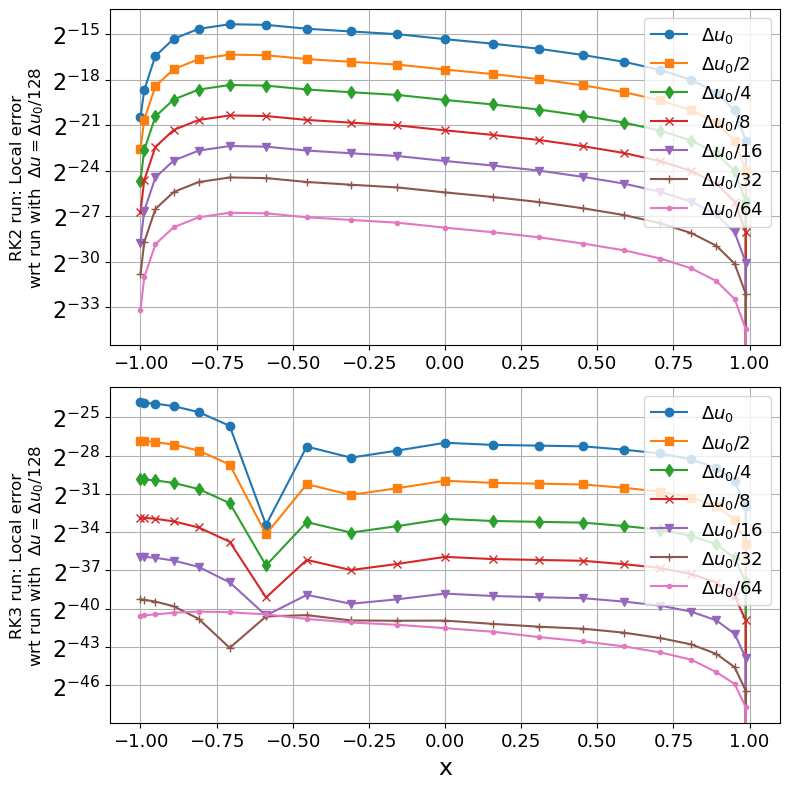}
\caption{\label{figConvTimeLocal} 
{Absolute errors of the profiles of ${\Phi(u=2.975,x)}$  when comparing low temporal resolution runs with ${\Delta u\in\{\Delta u_0, \hdots, \Delta u_0/64\}}$ to the high temporal resolution run with $
\Delta u_0/128$. }
} 
 \end{figure}
\begin{figure}[h]
\includegraphics[width=0.45\textwidth]{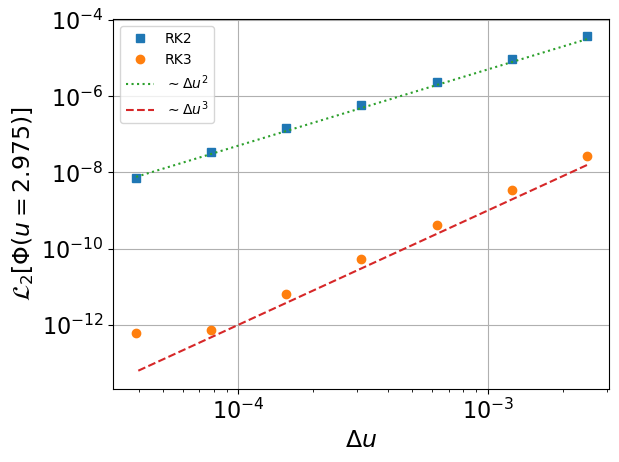}
\caption{\label{figConvTimeL2} 
{$\mathcal{L}_2$ norm of $\Phi$ as function of $\Delta u$ at the final time ${u=2.975}$. 
}
} 
 \end{figure}
  
An additional feature is the ability to track  the position $\lambda_A$ of the apparent horizon. If at  time $u_{[r_{,\lambda}=0]}$, $r_{,\lambda}$ changes sign, i.e. $r_{,\lambda}(u_{[r_{,\lambda}=0]}, \lambda_A) = 0$, the $\lambda-$grid can be automatically adopted such that $\lambda_A$ corresponds to the
outer boundary $x=1$ of the computational domain. For that purpose, the value  of $\lambda_A$  is determined by a standard Newton-Raphson method using the $\lambda-$position of the midpoint between the maximum and minimum of $r_{,\lambda}$ as initial guesses.  We then choose $A _{\lambda_A}= \lambda_A/2$ as the new parameter 
in the grid function \eqref{lambda_lin}. 

{Inside the event horizon, the spectral coordinate $x$
obeys $ \lambda=A(x+1)$,
where the outer boundary is
at $\lambda =2A$. Here $A$ must be chosen
so that $2A<\lambda_C$  so that the singularity
is excised from the computational domain. For this purpose, at each time step $u_n$ we choose the
value $A=A_n$ such that
$2A_n=\lambda_A(u_n)$, i.e. we locate the outer boundary
at the apparent horixon. We then integrate the hypersurface equations at time $u_n$. This determines $\Phi_{,u}(u_n)$
which allows the update to timestep $u_{n+1}$. This
is carried out with the value of $A=A_n$ fixed at time step $u_n$. We could, for some time, continue to evolve with this value $A=A_n$ since $\lambda_A(u)$ is a decreasing function and the
outer boundary would remain outside the apparent horizon.
However, 
eventually the outer boundary would approach the singularity.
In order to avoid this, after updating $\Phi(u_{n+1})$
and integrating the hypsurface equations, we find
$\lambda_{A}(u_{n+1})$ and determine the corresponding value
of $A_{n+1}$. This value is then used to set the outer boundary for evolution to the next timestep. Note that $A$ is set to
the constant value $A_n$ in
the evolution from $u_n$ to $u_{n+1}$. Thus the time dependence of
$A$ does not enter the integration of the evolution scheme. 
In turn, the data $\Phi(x_i)$ on the grid determined by $ A_n$ are mapped to the  grid  determined by $A_{n+1}$ using a cubic spline.}

Our numerical implementation reproduces the standard features of the critical solution such as the universality and echoing (Fig.~\ref{echoing}), scalar field dispersion (Fig.~\ref{dispersion}) and scalar field collapse (Fig.~\ref{collapes}). In these figures,  we used $a=0.7$ for the inital data \eqref{Phi_data}, for which $u^*\approx 2.1$.  

\begin{figure}[h]
\includegraphics[width=0.4\textwidth]{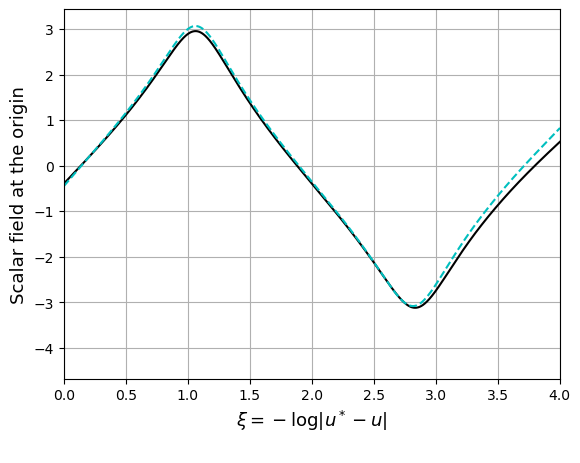}
\caption{\label{echoing}
The echoing and universality shown for two different  initial data sets: The black line corresponds to  the initial data of \cite{CW2019}, i.e. $\Phi_{[12]}(0,\lambda) = \epsilon(1+\lambda^2)^{-1}$ with $\epsilon=2.2731644$.  The cyan dashed line corresponds the initial data \eqref{Phi_data} with  ${a=0.7}$ and $b=0.947201675$. The echoing period is consistent with the Choptuik value $\Delta=3.44$, while the critical times are $u^*_{[12]} = 2.2039$  and $u^ *_{(3.4)}=2.165067$.
}
\end{figure}
\begin{figure}[h!]
\begin{center}
\includegraphics[width=0.4\textwidth]{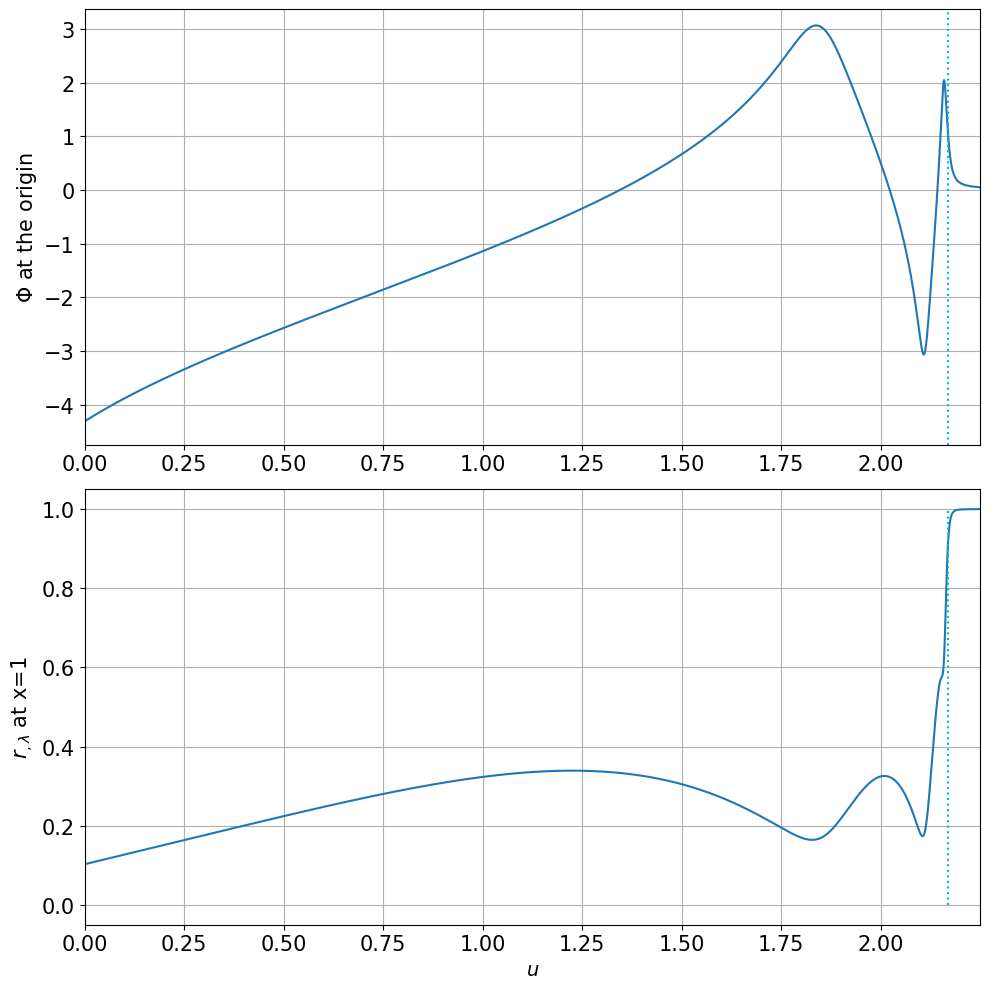}
\end{center}
\caption{\label{dispersion} Standard behavior of the critical solution: Subcritical initial data evolve toward the Minkowski value ${\lim_{\lambda\rightarrow\infty}r_{,\lambda} = H\rightarrow 1}$ and the  scalar field disperses, $\Phi\rightarrow 0$, at
time $u>u^*\approx2.1$ (vertical dotted line). }
\end{figure}

\begin{figure}[h!]
\begin{center}
\includegraphics[width=0.4\textwidth]{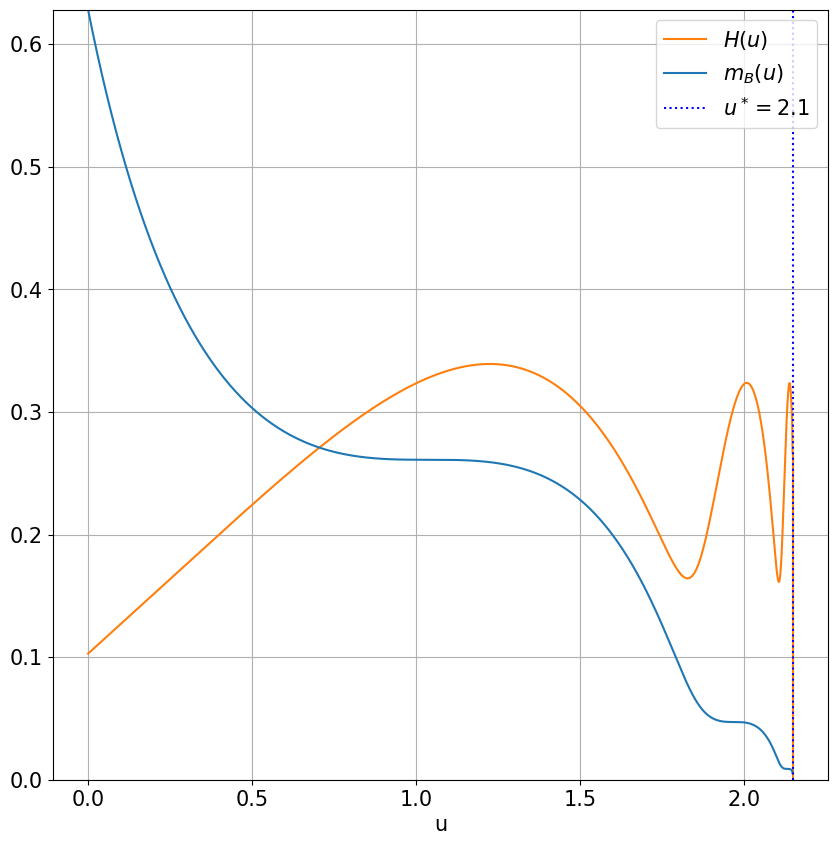}
\end{center}
\caption{\label{collapes} Standard behavior of the critical solutions: For supercritical initial data, the Bondi mass  decays to a finite value and  $H$ goes to zero for  $u\rightarrow u^*\approx 2.1$ (vertical dotted line). }
\end{figure}

\section{New results}

\subsubsection{Scale symmetry}
{Systematic investigation of the  2-parameter
space $(a,b)$ for the
initial data \eqref{Phi_data}  facilitates the study
of critical phenomena. The scale symmetry of the
initial data with respect to the parameter $a$  is
consistent with the scale symmetry of the evolution
system \eqref{EFE}. As a result, the value of $a$ sets an overall length
scale for the solution, e.g. the critical time $u^*$ depends linearly on $a$.
Numerical evolution determines the value $u^* \approx 3 a$.
This explicit knowledge allows an estimate of the number of time steps needed to study critical collapse.
We also find that the critical parameter $b^*\approx 0.947202$  for all values of $a$.  }

\subsubsection{Tracking the apparent horizon}
To demonstrate that the system \eqref{hier_alt} can track
the position of the apparent horizon, we use inside event horizon data with $b >1$.
Because the position $\lambda_A$
of the apparent horizon moves inward towards the central geodesic and because the $\lambda$ domain shrinks in the process, we adaptively remap $\lambda_A$ to the outer boundary $x=1$ by adjusting the grid parameter $A$. This   remapping is done after every time step and allows us to follow the motion of the apparent horizon close to the  central world line. 
Fig.~\ref{Fig:profiles} shows temporal snapshots of radial profiles of $r$, the Misner-Sharp mass $m$, $r_{,\lambda}$ and the negative Ricci scalar for the evolution of the same inside event horizon initial data presented in Fig.~\ref{iData}. The apparent horizon is initially at
$\lambda_A(u=0)=2.5(\sqrt{2}-1)\approx 1.04$.
We  follow the motion of the apparent horizon up until $\lambda_A(u=0.67) = 0.05$. At this final time,  the absolute value of the Ricci  scalar $|g^{ab}R_{ab}|$ has risen by an order of magnitude over its value on the initial slice. 

{As $\lambda_A(u)$ approaches the central world line, we confirm that the area of the apparent horizon decays. 
As expected from the relation  $r(u, \lambda_A) = 2m(u, \lambda_A)$, which follows from \eqref{def_M_ms},
the Misner-Sharp mass of the apparent horizon $m_A(u)$
also decays as the apparent horizon approaches the
central worldline. For simulations run successively with 101, 201, 401 and 801 grid points, numerical convergence was measured by comparing the positions $\lambda_A(u_n)$  of the low resolution runs ($N\in\{101, 201, 401\}$) with the  highest resolution run $N=801$. As shown in Fig.~\ref{Fig:decay_r_mms},
the respective errors differ by a factor of 4 throughout the  evolution, which confirms the convergence of the second order time integrator.} 

In Fig.~\ref{Fig:snapshots_drdl_r_ms}, the position of the apparent horizon $\lambda_A$ can be read off from where
$r_{,\lambda}=0$. 
Fig.~ \ref{Fig:snapshots_drdl_r_ms} also shows that the
profiles of $r$ and $2m$ intersect at the origin and then again at the position of  apparent horizon.  
\begin{figure}[h!]
\begin{center}
\includegraphics[width=0.4\textwidth]{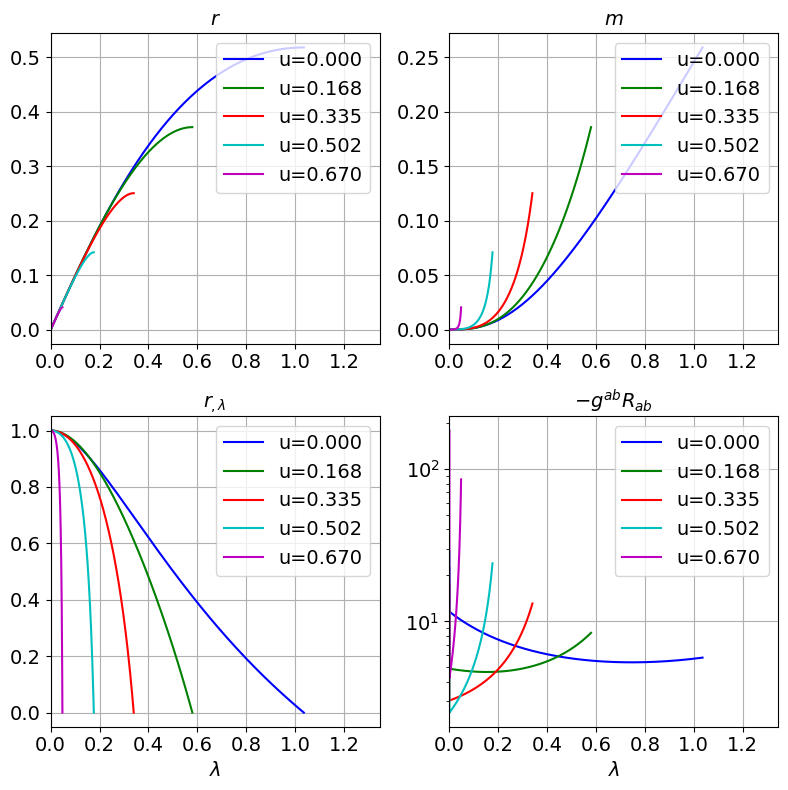}
\end{center}
\caption{\label{Fig:profiles}
Snapshots of profiles of areal distance $r$ (upper left),   Misner-Sharp mass $m$ (upper right), $r_{,\lambda}$ (lower left) and negative Ricci scalar (lower right) for inside event horizon initial data. Values at the endpoints of the curves  correspond to values at the apparent horizon at the  time indicated by the color in the legend.}
\end{figure}

\begin{figure}[h!]
\begin{center}
\includegraphics[width=0.4\textwidth]{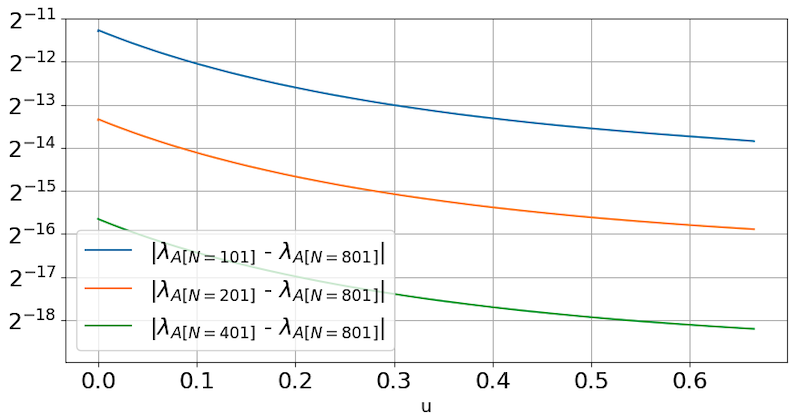}
\end{center}
\caption{\label{Fig:decay_r_mms} 
Comparison of the absolute error of the position of the apparent horizon
$\lambda_A(u)$ between  low resolution  runs  ($N\in\{101,201,401\}$)  and the high resolution run with $N=801$ grid points.}
\end{figure}

\begin{figure}[h!]
\begin{center}
\includegraphics[width=0.4\textwidth]{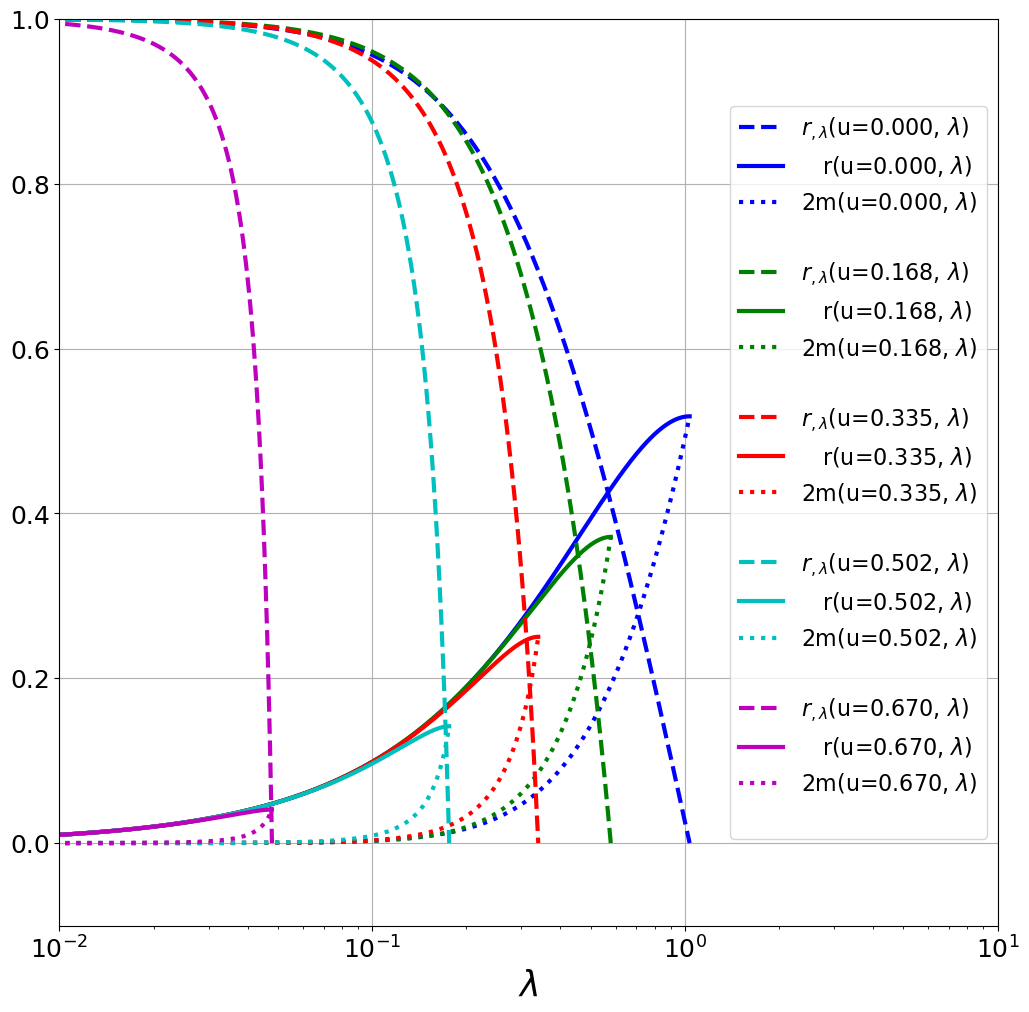}
\end{center}
\caption{\label{Fig:snapshots_drdl_r_ms}
Snapshots of profiles of areal distance $r$ (solid lines),   $r_{,\lambda}$ (dashed lines) and twice the Misner-Sharp mass $2m$ (dotted lines) between the origin and the apparent horizon. The profiles of  $r$ and $2m$ intersect at the apparent horizon. The  colors indicate the times of the snapshots
as indicated in the legend. 
}
\end{figure}

\subsubsection{Evolution across the event horizon}
For asymptotically flat supercritical initial data with $a=1$ and $b=0.95$, we are able to continue the evolution of  asymptotically flat data across the event horizon
up to the final collapse of the apparent horizon to the central worldline.
For these simulations, we use the compactified grid function \eqref{lambda_compact} with $A=1$ in the exterior region, where the $\lambda$ domain extends from the central world line to null infinity ${\cal I}^+$. 
At central time $u_H$, the time when the event horizon forms, $r$ approaches $2M_{B}$  as $\lambda$ goes to infinity.
We determine  $u_H\approx 2.027$ from the average of the time we detect a singular caustic on an outgoing null cone ($u_C=2.02813$) and the time of the last null cone that extends to ${\cal I}^+$ ($u=2.02687$).
For times $u>u_H$, we excise the singular caustic and follow the motion of the apparent horizon using the linear grid function \eqref{lambda_lin}. 
Fig.~\ref{Fig:grid_evolution} shows this adaptation of the grid function and grid parameter $A$ during the evolution of supercritical data.

\begin{figure}[h!]
\begin{center}
\includegraphics[width=0.4\textwidth]{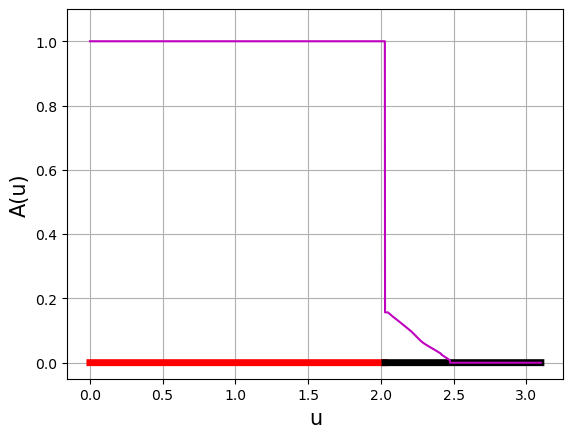}
\end{center}
\caption{\label{Fig:grid_evolution} The red line indicates the use of the compactified grid function in the  exterior of the event horizon and the black line the use of a linear grid function in the  interior of the event horizon, during a supercritical run when the event horizon forms at $u_H\approx2.027$. The magenta  line indicates the change of the grid parameter $A$ from its initial value $A=1$ for $u<u_H$ and its decay towards zero for $u>u_H$ due to the remapping of the apparent horizon. 
}
\end{figure}

\begin{figure}[h!]
\begin{center}
\includegraphics[width=0.4\textwidth]{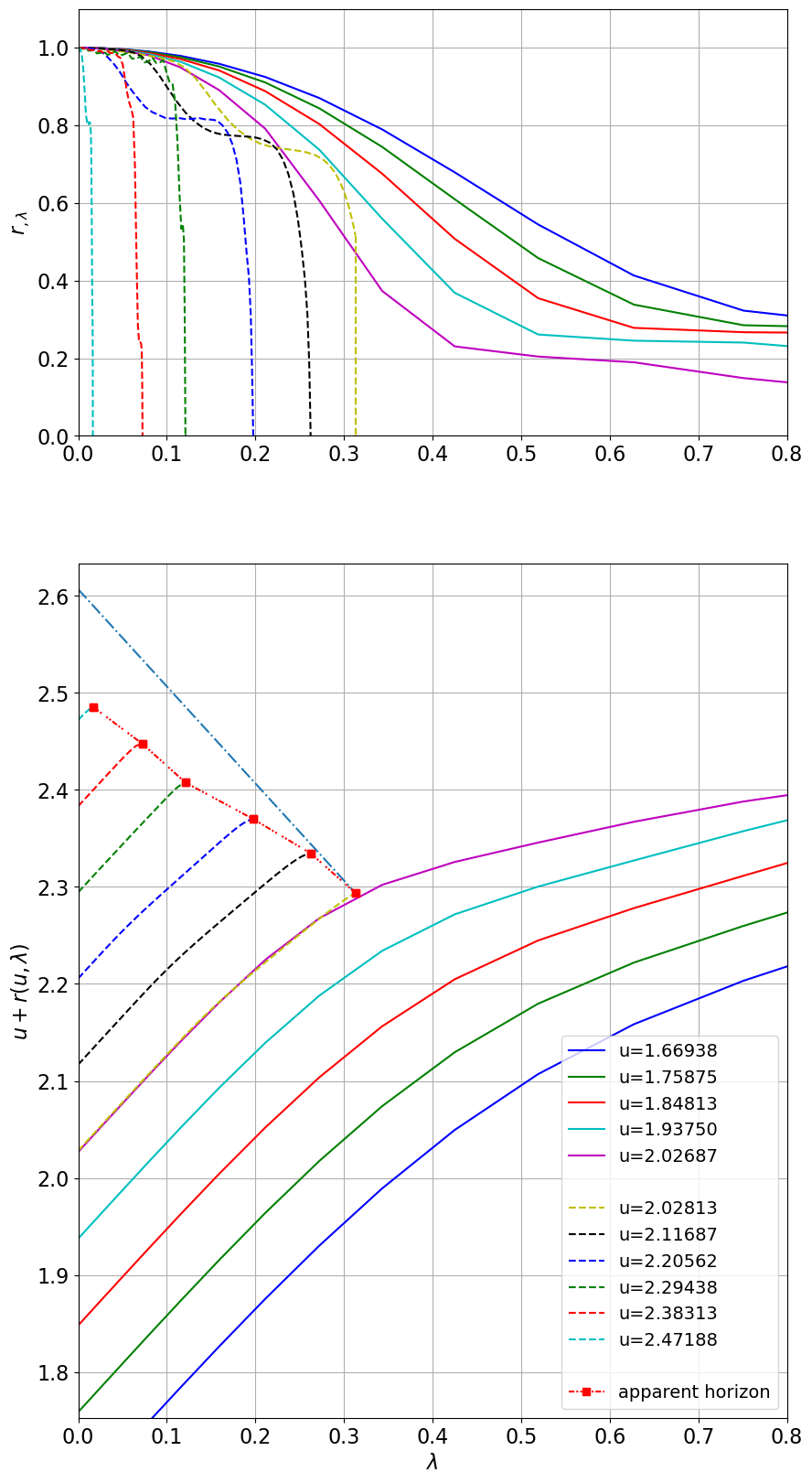}
\end{center}
\caption{\label{Fig:u+r}
Temporal snapshots for supercritical initial data  showing $r_{,\lambda}$ (top panel) and $u+r(u, \lambda_{A})$ (bottom panel) before and after event horizon formation
at $u_H\approx2.027$: In the lower panel, the locus of the apparent horizon (red dash dotted  curve)  lies below the tangent (blue dashed-dotted  line) to the ingoing null ray at the position of the first detected apparent horizon.
This is consistent with the spacelike nature of the
apparent horizon. The color coded times in the legend apply to both panels. }
\end{figure}

Fig.~\ref{Fig:u+r}, shows snapshots of the evolution in the domain $0\le \lambda\le0.8$.
The upper panel shows  profiles of  $r_{,\lambda}$ before and after event horizon formation at $u_H$.
In the lower panel, 
for cleaner visualisation of the snapshots of $r(u,\lambda)$, we instead plot $u+r(u, \lambda)$ before and after event horizon formation. (Snapshots of $u+r$ rather than $r$ separate the profiles near the origin.)
In both panels of Fig.~\ref{Fig:u+r}, the solid lines are profiles in the exterior of the event horizon and the dashed lines in the interior, while the same colors correspond to profiles at
the same times.

In the upper panel,  $r_{,\lambda}$ decays for $u<u_H$, 
consistent with the increasing
redshift between an exterior observer
and the  central worldline.
For values $u\rightarrow u_H$ in  Fig.~\ref{Fig:u+r}, we see an inward motion of the positions on the $\lambda$-axis  where ${r_{,\lambda}=0}$, i.e. the positions $\lambda_A(u)$
of the apparent horizon. 
In the lower panel of Fig.~\ref{Fig:u+r}, the corresponding values of $u+r(u,\lambda_A)$ are depicted by red squares. 
The dashed-dotted line connecting these squares indicate the locus of the apparent horizon. The blue dashed-dotted line Fig.~\ref{Fig:u+r} at $(\lambda, u+r) \approx (0.31, 2.29)$ is the tangent to the ingoing null ray emanating from the position of the first detected apparent horizon.
The locus of the apparent horizon   lies  below this ingoing null direction, consistent with the spacelike
nature of the apparent horizon hypersurface.

In the lower panel of Fig.~\ref{Fig:u+r}, all profiles of
$u+r$ start from the center with the same slope
$r_{,\lambda}(u,0)=1$, as required by the local
Minkowski coordinate conditions at the origin. As $\lambda$
increases, the profiles deviate from straight lines and
become concave due to the focusing of the null rays by the scalar field. 

\section{Summary}
The Choptuik critical phenomena is a pristine problem in dynamic black hole formation. Here, we re-investigated the gravitational collapse of a massless scalar field in spherical symmetry using a characteristic formulation.
In comparison with other studies \cite{2002PhRvD..65h1501L,Purrer:2004nq}, which used the  Bondi-Sachs metric  \cite{MW2016}, we employed an affine-null metric. This allows numerical evolution beyond the formation of an event horizon where the areal coordinate $r$ in the Bondi-Sachs
formalism becomes singular. An obstacle 
in implementing the affine-null formulation is that
the main equations do not form a simple hierarchical scheme,
as in the Bondi-Sachs formalism. However, the hierarchical structure
has been restored  for the  general vacuum Einstein equations using a change in evolution variables \cite{Win2013,TM2019}, and this has been applied to the spherically symmetric Einstein-scalar field equations \cite{CW2019}. We adopted these variables here in the hierarchical system \eqref{hier1}. However, although the affine-null coordinates remain non-singular up to the formation of physical singularities, 
there are individual terms in the system \eqref{hier1} which are
infinite at the location of an apparent horizon, where
$1/r_{,\lambda}\rightarrow \infty$.
This is a limitation for applications of \eqref{hier1}
to evolution in the interior of a black hole (as studied in \cite{Win2013,TM2019}).

However,  as shown in \cite{CW2019}, in spherical symmetry it is possible to regularize the system \eqref{hier1} such that it is free of the troublesome $1/r_{,\lambda}$ terms. This allows simulations of gravitational collapse to penetrate the event horizon. Here we took
further advantage of this approach to
implement an independent version of this regularized hierarchical system to explore the dynamics
of the apparent horizon.

We verified the main results of \cite{CW2019} using a different pseudo-spectral method.
In addition, we have shown that supercritical initial data could be evolved beyond the event horizon to the interior of the black hole (see Fig.~\ref{Fig:u+r}). This  evolution could be followed almost to the
final singularity when the
area of the apparent horizon approaches zero. The evolution of
supercritical data demonstrated that the apparent horizon is a space-like hypersurface,
in accord with analytic results. Analytic results
using the affine-null system also showed that the final
singularity is a space-like hypersurface. 
Our results led to the space-time picture Fig.~\ref{conf_supercritical} of supercritical gravitational collapse.

We also presented new null cone initial data \eqref{Phi_data} which is well suited for investigating both sub-critical and super-critical evolutions. For this initial data the hierarchy of hypersurface equations could be integrated to yield
all metric and auxiliary functions in
closed analytic form (see \eqref{ana_idata}).
It is natural to ask if a regularized hierarchy such as \eqref{hier_alt} can be found for systems with less symmetry or different matter sources. A conclusive  answer to this  question is not in sight but  under investigation.

\section{Acknowledgements}
TM was supported by the FONDECYT de iniciaci\'on, 2019, No. 11190854.
OB  acknowledges a PhD scholarship of the University of Talca. HH is supported by the ANID PhD  fellowship No. 21310374 of the Chilean government.
\bibliographystyle{ieeetr}

\bibliography{ref.bib}

\end{document}